\documentclass[final,5p,twocolumn,times,sort&compress]{elsarticle}

\usepackage{pdfsync}
\usepackage{mathrsfs}
\usepackage[latin1]{inputenc}
\usepackage{array}
\usepackage{dsfont}
\usepackage{amsmath}\usepackage{amssymb,stmaryrd}
\usepackage{booktabs}
\usepackage{lipsum}
\usepackage{cancel}
\usepackage{xcolor}
\usepackage{hyperref}
\usepackage{widetext}

\def\preprintdate{February 2021}

\renewcommand{\eqref}[1]{\mbox{Eq.~(\ref{#1})}}

\newcommand{\secref}[1]{\mbox{Sec.~\ref{#1}}}

\begin{document}

\begin{frontmatter}

\title{Classical Lagrangians for the nonminimal spin-nondegenerate \\ Standard-Model Extension at higher orders in Lorentz violation}

\author[ufma,uema]{Jo\~{a}o A.A.S Reis}\ead{joao.andrade@discente.ufma.br}
\author[ufma]{Marco Schreck\corref{cor1}}\ead{marco.schreck@ufma.br}
\cortext[cor1]{Corresponding author}

\address[ufma]{Departamento de F\'{i}sica, Universidade Federal do Maranh\~{a}o \\
Campus Universit\'{a}rio do Bacanga, S\~{a}o Lu\'{i}s -- MA, 65085-580, Brazil}
\address[uema]{Departamento de F\'{i}sica, Centro de Educa\c{c}\~{a}o, Ci\^{e}ncias Exatas e Naturais \\
Universidade Estadual do Maranh\~{a}o, Cidade Universit\'{a}ria Paulo VI, S\~{a}o Lu\'{i}s -- MA, 65055-310, Brazil}

\address{\rm
\preprintdate
}

\begin{abstract}

We present new results for classical-particle propagation subject to Lorentz violation. Our analysis is dedicated to spin-nondegenerate operators of arbitrary mass dimension provided by the fermion sector of the Standard-Model Extension. In particular, classical Lagrangians are obtained for the operators $\hat{b}_{\mu}$ and $\hat{H}_{\mu\nu}$ as perturbative expansions in Lorentz violation. The functional dependence of the higher-order contributions in the background fields is found to be quite peculiar, which is probably attributed to particle spin playing an essential role for these cases. This paper closes one of the last gaps in understanding classical-particle propagation in the presence of Lorentz violation. Lagrangians of the kind presented will turn out to be valuable for describing particle propagation in curved backgrounds with diffeomorphism invariance and/or local Lorentz symmetry explicitly violated.

\end{abstract}


\end{frontmatter}

\section{Introduction}

Theories of physics at the Planck scale such as strings~\cite{Kostelecky:1988zi}, loop quantum gravity~\cite{Gambini:1998it}, noncommutative spacetime structures~\cite{AmelinoCamelia:1999pm}, and spacetime foam~\cite{Klinkhamer:2003ec} as well as nontrivial spacetime topologies~\cite{Klinkhamer:1998fa} and Ho\v{r}ava-Lifshitz gravity~\cite{Horava:2009uw} predict violations of Lorentz invariance. Planck-scale physics is presently not testable by direct means, but the low-energy fingerprints of Lorentz violation could show up in feasible experiments performed at much lower energies. To be able to translate the absence of signals for Lorentz violation into constraints on meaningful physical quantities, the Standard-Model Extension (SME) was constructed~\cite{Colladay:1996iz,Kostelecky:2003fs} as an effective field theory framework that parameterizes deviations from Lorentz symmetry.

The SME can be decomposed into a nongravitational~\cite{Colladay:1996iz,Kostelecky:2009zp,Kostelecky:2013rta,Kostelecky:2018yfa} and a gravitational part~\cite{Kostelecky:2003fs,Kostelecky:2020hbb} where each of them on its own consists of a minimal and a nonminimal sector. The minimal SME incorporates Lorentz-violating field operators of mass dimensions 3 and 4, whereas the nonminimal SME comprises all higher-dimensional operators. The nonminimal contributions for photons, neutrinos, and Dirac fermions without gravity are neatly classified in~\cite{Kostelecky:2009zp,Kostelecky:2013rta} where \cite{Kostelecky:2018yfa} provides a more general classification that respects the gauge structure of the Standard Model (SM). The most recent work \cite{Kostelecky:2020hbb} clarifies various aspects of Lorentz and diffeomorphism violation in gravity and introduces a method to construct operators of arbitrary mass dimension that are invariant under general coordinate transformations. At large, investigations of Lorentz violation in gravity can be intricate, which is why linearized gravity is the base of a series of papers such as~\cite{Bailey:2014bta,Kostelecky:2016kfm}. Searches for {\em CPT}-violation are sometimes considered as more important than searches for Lorentz violation. Since {\em CPT} violation implies Lorentz violation in effective field theory according to a theorem by Greenberg~\cite{Greenberg:2002uu}, the SME automatically comprises all {\em CPT}-violating operators.

One of the great successes of the SME over the past two decades is the large number of tight constraints on both minimal and nonminimal Lorentz violation in the nongravitational part~\cite{Kostelecky:2008ts}. Lorentz violation in the gravity sector has also been constrained, but most of these bounds are significantly weaker than those of the nongravitational SME. The reason is undoubtedly that gravitational experiments of high precision are more challenging to perform in comparison to experiments that are insensitive to gravity. After all, the experimental value of Newton's constant has the largest experimental uncertainty in comparison to the remaining constants of nature.

Therefore, Lorentz violation may well hide in the gravitational sector and may have remained unnoticed, so far. Also, there are certain types of Lorentz violation that cannot be observed in Minkowski spacetime even if they are enormous, as they can be removed from the Lagrange density by a redefinition of the physical fields. However, such a redefinition loses its validity in the presence of gravity, whereupon the enormous value would be suppressed by the weakness of the gravitational interaction~\cite{Kostelecky:2008in}.

Earthbound experiments searching for deviations from the laws of General Relativity include tests of the weak equivalence principle performed with test masses of different materials in drop-tower experiments as well as torsion pendulum and gravimeter experiments. Furthermore, space-born experiments such as Gravity Probe B or laser ranging experiments can set bounds on violations of local Lorentz invariance in gravity (see~\cite{Bailey:2006fd} for a compilation of such tests). Lorentz violation in the gravity sector can also be constrained via the absence of gravitational vacuum Cherenkov radiation that would imply energy losses of ultra-high-energetic particles by the emission of gravitons~\cite{Moore:2001bv}. The announcement of the direct detection of gravitational waves in 2016~\cite{Abbott:2016blz} opened another window for searches for Lorentz violation in gravity, as the latter would modify the emission and propagation properties of gravitational waves~\cite{Kostelecky:2016kfm,Xu:2021dcw}.

Earth-based experiments in the gravitational sector involve extended test masses that are beyond the field-theory description of the SME. Hence, to test local Lorentz invariance in gravity, it is desirable work with an SME-equivalent that parameterizes Lorentz violation for classical, relativistic, pointlike particles. An algorithm that provides such a description was proposed in~\cite{Kostelecky:2010hs} and since then it has been applied to various sets of Lorentz-violating operators~\cite{LagFinslerPapers,Schreck:2014hga,Schreck:2015seb,Reis:2017ayl,Edwards:2018lsn,Schreck:2019mmr}. Some of these classical Lagrangians are known to be connected to Finsler structures~\cite{FinslerPapers} that give rise to Finsler geometries. Finsler geometry is less restrictive than Riemannian geometry \cite{Riemann:2004,Darrigol:2015}, as it does not necessarily rely on quadratic path length functionals~\cite{Finsler:1918,Antonelli:1993,Bao:2000}.

Further studies in this realm involve the reverse process from a Lagrangian back to the Hamiltonian of a field theory~\cite{Schreck:2016jqn}, methods of removing singularities of Finsler spaces~\cite{Colladay:2015wra} according to Hironaka's theorem \cite{Hironaka:1964}, applications within classical mechanics and electrostatics~\cite{Foster:2015yta} as well as investigations of scalar fields in Finsler spaces~\cite{Silva:2015ptj}. Results on Finsler structures that occur for Lorentz-violating photons in the eikonal limit are also available~\cite{Schreck:2015dsa}.

To complete the picture, a systematic treatment of the spin-nondegenerate fermion operators is still missing. Although results for nonminimal operators at first order in Lorentz violation are already available \cite{Reis:2017ayl}, nothing is known about the structure of higher-order contributions. This gap shall be closed with the current article.

Our paper is organized as follows. Section~\ref{sec:dirac-theory} gives a summary on the most important characteristics of the modified Dirac fermion sector of the nonminimal SME that will be of importance in the subsequent analysis. Section~\ref{sec:classical-kinematics} summarizes the procedure of how to map a field theory to the description of a classical, relativistic, pointlike particle in terms of a Lagrangian. In \secref{sec:results-perturbative-expansion} some intermediate results are obtained that allow us to determine the perturbative expansion of the Lagrangians considered. An application of the method in \secref{sec:third-order-lagrangians} provides the covariant Lagrangian for dimension-5 $b$ and $H$ coefficients at second order in Lorentz violation. Finally, all findings are concluded on in \secref{sec:conclusions}. The appendix presents worthwhile computational details that are of lesser interest to be shown in the main body of the paper. Natural units will be used with $\hbar=c=1$ unless otherwise stated.

\section{Dirac field theory modified by spin-nondegenerate operators}
\label{sec:dirac-theory}

The Dirac fermion sector of the SME describes modified spin-1/2 fermions that are subject to Lorentz violation. Its nonminimal version including the full spectrum of possible Lorentz-violating operators was constructed in \cite{Kostelecky:2013rta} and it has the form
\begin{equation}
\label{eq:sme-fermion-sector}
\mathcal{L}=\frac{1}{2}\overline{\psi}(\mathrm{i}\cancel{\partial}-m_{\psi}\mathds{1}_4+\hat{\mathcal{Q}})\psi+\text{H.c.}\,,
\end{equation}
where $\psi$ is a Dirac spinor field, $\overline{\psi}\equiv \psi^{\dagger}\gamma^0$ the Dirac-conjugated field, $m_{\psi}$ the fermion mass, and $\mathds{1}_4$ the identity matrix in spinor space. All fields are defined in Minkowski spacetime with metric $\eta_{\mu\nu}$ of signature $(+,-,-,-)$. Furthermore, $\cancel{\partial}\equiv \gamma^{\mu}\partial_{\mu}$ with the standard Dirac matrices satisfying the Clifford algebra $\{\gamma^{\mu},\gamma^{\nu}\}=2\eta^{\mu\nu}\mathds{1}_4$.

The operator $\hat{\mathcal{Q}}$ comprises all contributions that are in accordance with the spinor structure of Dirac theory. It consists of a spin-degenerate part that involves the operators $\hat{a}_{\mu}$, $\hat{c}_{\mu}$, $\hat{e}$, $\hat{f}$, and $\hat{m}$. Spin-degenerate Lorentz violation does not lift the twofold degeneracy of the fermion dispersion relation, i.e., the energy-momentum dependence for spin-up fermions is the same as that for spin-down fermions. Apart from these operators, $\hat{\mathcal{Q}}$ contains a spin-nondegenerate part including the operators $\hat{b}_{\mu}$, $\hat{d}_{\mu}$, $\hat{H}_{\mu\nu}$, and $\hat{g}_{\mu\nu}$. Dirac fermions that interact with background fields of these types have propagation properties dependent on their spin projection. In other words, the dispersion relation of spin-up fermions differs from that for spin-down fermions. Each of these operators mentioned can be decomposed into a sum of controlling coefficients contracted with a certain number of four-derivatives that successively increases by 2. The mass dimension $d$ of these controlling coefficients depends on the number of four-derivatives that occur in the operator. Coefficients of negative mass dimensions are contained in the nonminimal SME and the corresponding operators are power-counting nonrenormalizable.

In general, the structure of the spin-nondegenerate operators is way more involved than that of the spin-degenerate ones, which is a property that will also become evident in the forthcoming sections of the paper. The underlying reason is their dependence on the spin projection, which is absent for the spin-degenerate operators. For example, the complicated structure of the spin-nondegenerate operators shows up in their dispersion relations, the spinor solutions of the modified Dirac equation~\cite{Kostelecky:2000mm,Kostelecky:2013rta,Reis:2016hzu} as well as the plethora of different behaviors in unusual particle physics processes such as vacuum Cherenkov radiation~\cite{Schreck:2017isa}.

\section{Classical kinematics}
\label{sec:classical-kinematics}

To evaluate data from experimental tests of local Lorentz invariance in gravity, it is desirable to have a framework parameterizing Lorentz violation for classical, pointlike particles. First of all, gravity is dominant for macroscopic test bodies, whereupon their behavior is dominated by the laws of classical physics. Second, with the law of motion of a pointlike particle at hand, at least the translational behavior of a macroscopic object consisting of these particles is obtained from integrations over suitable mass distributions.

The SME itself is based on field theory and does not describe Lorentz violation for classical particles. Therefore, the modified Dirac fermion sector of \eqref{eq:sme-fermion-sector} must be mapped suitably to the Lagrangian $L=L(u)$ of a relativistic pointlike particle moving with four-velocity $u^{\mu}$. A reasonable map was constructed around ten years ago in \cite{Kostelecky:2010hs} and it is governed by a set of five ordinary, nonlinear equations:
\begin{subequations}
\label{eq:mapping-equations}
\begin{align}
\label{eq:mapping-equation-1}
\mathcal{D}(p)&=0\,, \displaybreak[0]\\[1ex]
\label{eq:mapping-equation-2-3-4}
\frac{\partial p_0}{\partial p_i}&=-\frac{u^i}{u^0}\,,\quad i\in \{1,2,3\}\,, \displaybreak[0]\\[1ex]
\label{eq:mapping-equation-5}
L&=-p_{\mu}u^{\mu}\,.
\end{align}
\end{subequations}
Equation~(\ref{eq:mapping-equation-1}) is the dispersion equation of the modified Dirac theory in \eqref{eq:sme-fermion-sector}. The latter depends on the four-momentum $p_{\mu}$ and follows from the requirement that the modified Dirac equation have nontrivial spinor solutions. Equations~(\ref{eq:mapping-equation-2-3-4}) say that the centroid of a wave packet constructed from plane-wave solutions of the modified Dirac equation moves with a group velocity equal to the three-velocity $\mathbf{u}/u^0$ of the corresponding classical particle. The minus sign on the right-hand side takes into account the different positions of the spatial index $i$ on both sides of the equations. Finally, the Euler equation~(\ref{eq:mapping-equation-5}) holds for a Lagrangian $L$ that is positively homogeneous of degree~1: $L(\lambda u)=\lambda L(u)$ for $\lambda>0$ (see \cite{Bao:2000}). The latter property is appealing for physical reasons, as it implies an action that is invariant with respect to reparameterizations of the classical trajectory.

The four-momentum in the dispersion equation is interpreted as the canonical momentum connected to the Lagrangian via
\begin{equation}
p_{\mu}=-\frac{\partial L}{\partial u^{\mu}}\,.
\end{equation}
The five equations~(\ref{eq:mapping-equations}) depend on the four-momentum components, the four-velocity components, and the classical Lagrangian to be determined. They should allow us to express the four-momentum completely in terms of the four-velocity and to state the Lagrangian as a function of the four-velocity. However, what is expected to work in theory, is challenging in practice, since the equations are both nonlinear and coupled. Over the past decade, different techniques were applied to successfully obtain Lagrangians for various sets of controlling coefficients. The first Lagrangians were derived in \cite{Kostelecky:2010hs} probably by directly manipulating the system of equations. By doing so, the authors obtained Lagrangians for both the spin-degenerate and the spin-nondegenerate fermion sector of the minimal SME. Further results followed in \cite{LagFinslerPapers}.

It was soon realized that classical Lagrangians in the context of the nonminimal SME were more challenging to derive. Nonminimal operators come with additional powers of four-momentum components increasing the nonlinearity of Eqs.~(\ref{eq:mapping-equations}). It seemed that Lagrangians exact in Lorentz violation may be highly nontransparent and too unwieldy to be used in applications \cite{Schreck:2014hga}. Therefore, the focus changed to obtaining such Lagrangians at first order in Lorentz violation only. This was done to simplify computations, but such results were also thought to be sufficient from a practical viewpoint, as Lorentz violation (at least in Minkowski spacetime) is already tightly constrained~\cite{Kostelecky:2008ts}.

The first Lagrangians of the nonminimal SME at leading order in Lorentz violation were derived in \cite{Schreck:2015seb} by solving Eqs.~(\ref{eq:mapping-equations}) with the help of Gr\"{o}bner bases. Due to the complexity of spin-nondegenerate operators, the latter analysis only involves classical Lagrangians for spin-degenerate operators. Round about two years later, these findings were complemented in~\cite{Reis:2017ayl} by including spin-nondegenerate operators, which provided the full classical-particle equivalent to the SME at leading order in Lorentz violation. The experience on the form of such Lagrangians gained in the years before greatly contributed to obtaining these results. Finally, a powerful method to derive perturbative series of Lagrangians in the Lorentz-violating coefficients was presented in~\cite{Edwards:2018lsn} and applied to a modified scalar field theory to obtain such series to third order in Lorentz violation. A subsequent analysis~\cite{Schreck:2019mmr} led to equivalent results for the spin-degenerate operators of the nonminimal Dirac fermion sector. So far, an analogous investigation for the spin-nondegenerate operators has not been carried out.

The intention of the current paper is to establish ties to the spin-nondegenerate operators. We will restrict the analysis to controlling coefficients $b^{(d)\mu\alpha_1\dots\alpha_{(d-3)}}$, $d^{(d)\mu\alpha_1\dots\alpha_{(d-3)}}$ that are totally symmetric and $H^{(d)\mu\nu\alpha_1\dots\alpha_{(d-3)}}$, $g^{(d)\mu\nu\alpha_1\dots\alpha_{(d-3)}}$ that are antisymmetric in the first two indices and totally symmetric in the remaining ones. Similar assumptions were taken to obtain perturbative series of Lagrangians for the spin-degenerate operators in~\cite{Schreck:2019mmr}. These restrictions are minor in comparison to how they simplify the computations. In many cases the coefficients with the described symmetries are dominant, whereas the others are suppressed (compare to the leading-order results of \cite{Reis:2017ayl,Edwards:2018lsn} that only involve the totally symmetric sets of coefficients).

The perturbative method first proposed in~\cite{Edwards:2018lsn} and applied to the spin-degenerate fermion operators in~\cite{Schreck:2019mmr} shall now be adopted suitably such that it can lead to perturbative series of Lagrangians for the spin-nondegenerate operators $\hat{b}_{\mu}$, $\hat{d}_{\mu}$, $\hat{H}_{\mu\nu}$, and $\hat{g}_{\mu\nu}$. To avoid couplings between different types of coefficients or coefficients of different mass dimensions, we only consider a particular coefficient type and a fixed mass dimension at a time. The crucial difference to the spin-degenerate operators is that the dispersion equation (even for the minimal framework) is no longer quadratic in the four-momentum, but quartic, at least. For $\hat{b}_{\mu}$ and $\hat{H}_{\mu\nu}$ they can be cast into the form
\begin{subequations}
\begin{align}
\label{eq:dispersion-equation-b}
\left|p^2-b^{(d)\mu\diamond}(b^{(d)})_{\mu}^{\phantom{\mu}\diamond}-m_{\psi}^2\right|&=2\Upsilon_b\,, \displaybreak[0]\\[2ex]
\label{eq:dispersion-equation-H}
|p^2+2\hat{X}-m_{\psi}^2|&=2\Upsilon_H\,,
\end{align}
with the valuable observer Lorentz scalars
\begin{align}
\Upsilon_b&\equiv\sqrt{(b^{(d)\diamond})^2-b^{(d)\mu\diamond}(b^{(d)})_{\mu}^{\phantom{\mu}\diamond}p^2}\,, \displaybreak[0]\\[2ex]
\Upsilon_H&\equiv\sqrt{2\hat{X}p^2-H^{(d)\nu\diamond}(H^{(d)})_{\nu}^{\phantom{\nu}\diamond}-\hat{Y}^2}\,, \displaybreak[0]\\[2ex]
\label{eq:definition-X-Y}
\hat{X}&\equiv\frac{1}{4}H^{(d)\mu\nu\diamond}(H^{(d)})_{\mu\nu}^{\phantom{\mu\nu}\diamond}\,,\quad \hat{Y}\equiv\frac{1}{4}H^{(d)\mu\nu\diamond}(\tilde{H}^{(d)})_{\mu\nu}^{\phantom{\mu\nu}\diamond}\,,
\end{align}
and the dual of $\hat{H}_{\mu\nu}$ in momentum space:
\begin{equation}
\tilde{H}^{(d)\mu\nu\diamond}\equiv\frac{1}{2}\varepsilon^{\mu\nu\varrho\sigma}(H^{(d)})_{\varrho\sigma}^{\phantom{\varrho\sigma}\diamond}\,.
\end{equation}
\end{subequations}
Here, $\varepsilon^{\mu\nu\varrho\sigma}$ is the totally antisymmetric Levi-Civita symbol in four spacetime dimensions with $\varepsilon^{0123}=1$. We employ the $\diamond$ notation that was introduced in \cite{Schreck:2019mmr} for convenience. It indicates coefficients suitably contracted with four-momenta (as opposed to four-velocities).

Comparing Eqs.~(\ref{eq:dispersion-equation-b}) and (\ref{eq:dispersion-equation-H}) with each other, it is evident that they share certain similarities, but there are also crucial differences. The observer scalar $\hat{Y}$ is nonzero only when there is at least one nonzero purely spatial component operator $\hat{H}_{ij}$ and a nonzero mixed one $\hat{H}_{0i}$. If $\hat{Y}=0$, we can directly identify
\begin{equation}
(b^{(d)\diamond})^2\leftrightarrow -H^{(d)\nu\diamond}(H^{(d)})_{\nu}^{\phantom{\nu}\diamond}\,,\quad b^{(d)\mu\diamond}(b^{(d)})_{\mu}^{\phantom{\mu}\diamond} \leftrightarrow -2\hat{X}\,,
\end{equation}
which reveals that there is a certain correspondence between these operators at the level of the dispersion equation.

\section{Basic results for perturbative expansion}
\label{sec:results-perturbative-expansion}

The procedure to obtain a perturbative expansion for a classical Lagrangian starts with a computation of the implicit derivative of the dispersion equation with respect to
$p_{\mu}$ and to use \eqref{eq:mapping-equation-2-3-4}. A contraction of the result with the spatial momentum components $p_j$ and a subsequent application of \eqref{eq:mapping-equation-5} as well as taking into account the general form of $\hat{b}_{\mu}$, $\hat{d}_{\mu}$, $\hat{H}_{\mu\nu}$, and $\hat{g}_{\mu\nu}$ for a fixed mass dimension $d$ implies the following four-velocities as functions of the four-momentum:

\begin{widetext}
\begin{subequations}
\label{eq:four-velocities}
\begin{align}
\label{eq:four-velocity-b}
u^{\mu}|_b&=-L\frac{\pm\Upsilon_b\left[p^{\mu}-(d-3)b^{(d)\mu\nu\diamond}(b^{(d)})_{\nu}^{\phantom{\nu}\diamond}\right]+b^{(d)\nu\diamond}(b^{(d)})_{\nu}^{\phantom{\nu}\diamond}p^{\mu}-b^{(d)\diamond}b^{(d)\mu\diamond}+(d-3)\left[b^{(d)\mu\nu\diamond}(b^{(d)})_{\nu}^{\phantom{\nu}\diamond}p^2-b^{(d)\diamond}b^{(d)\mu\diamond}\right]}{\pm\Upsilon_b\left[p^2-(d-3)b^{(d)\rho\diamond}(b^{(d)})_{\rho}^{\phantom{\rho}\diamond}\right]-(d-2)\Upsilon_b^2}\,, \displaybreak[0]\\[2ex]
\label{eq:four-velocity-H}
u^{\mu}|_H&=-L\frac{\pm \Upsilon_H[p^{\mu}+2(d-3)\hat{X}^{\mu}]-2\hat{X}p^{\mu}-H^{(d)\mu\nu\diamond}(H^{(d)})_{\nu}^{\phantom{\nu}\diamond}+(d-3)(-2\hat{X}^{\mu}p^2-p_{\nu}H^{(d)\nu\varrho\mu\diamond}(H^{(d)})_{\varrho}^{\phantom{\nu}\diamond}+2\hat{Y}\hat{Y}^{\mu})}{\pm \Upsilon_H[p^2+2(d-3)\hat{X}]-(d-2)\Upsilon_H^2+(d-4)\hat{Y}^2}\,,
\end{align}
\end{subequations}
\end{widetext}
\noindent where, for convenience, we additionally defined the useful observer four-vectors
\begin{equation}
\label{eq:vector-valued-X-Y}
\hat{X}^{\varrho}\equiv \frac{1}{4}H^{(d)\mu\nu\varrho\diamond}(\hat{H}^{(d)})_{\mu\nu}^{\phantom{\mu\nu}\diamond}\,,\quad \hat{Y}^{\varrho}\equiv \frac{1}{4}H^{(d)\mu\nu\varrho\diamond}(\tilde{H}^{(d)})_{\mu\nu}^{\phantom{\mu\nu}\diamond}\,.
\end{equation}
Calculational details on how to obtain these results explicitly are relegated to \ref{sec:computation-four-velocities}. Analogous results for $\hat{d}_{\mu}$ and $\hat{g}_{\mu\nu}$ follow from Eqs.~(\ref{eq:four-velocities}) via the replacements $b^{(d)\mu\diamond} \mapsto d^{(d+1)\mu\diamond}$ and $H^{(d)\mu\nu\diamond}\mapsto g^{(d+1)\mu\nu\diamond}$, respectively, and suitable adaptations of the mass dimension $d$. However, in contrast to results at first order in Lorentz violation (cf.~\cite{Reis:2017ayl}), it is not possible to perform the replacements $\hat{b}_{\mu}\mapsto -\hat{\mathcal{A}}_{\mu}$ and $\hat{H}_{\mu\nu}\mapsto -\hat{\mathcal{T}}_{\mu\nu}$ with the pseudoscalar and two-tensor operators $\hat{\mathcal{A}}_{\mu}$ and $\hat{\mathcal{T}}_{\mu\nu}$, respectively, of Eq.~(7) in \cite{Kostelecky:2013rta}. Such replacements would induce couplings between different types of coefficients and Eqs.~(\ref{eq:four-velocities}) are not valid for such scenarios.

Contracting a four-velocity of Eqs.~(\ref{eq:four-velocities}) with $u_{\mu}$ and using \eqref{eq:mapping-equation-5} again provides quadratic equations in the corresponding Lagrangians, such as for the case of spin-degenerate operators~\cite{Edwards:2018lsn,Schreck:2019mmr}. This finding is interesting, since the dispersion equations for the spin-nondegenerate operators are quartic at least (as mentioned before). For example, for the case of $\hat{b}_{\mu}$ we obtain
\begin{subequations}
\begin{equation}
\label{eq:equation-lagrangian}
0=\zeta_b^{\pm}L_b^2+\psi_b^{\pm}L_b-u^2\,,
\end{equation}
with
\begin{equation}
\zeta_b^{\pm}=\frac{\pm\Upsilon_b+b^{(d)\mu\diamond}(b^{(d)})_{\mu}^{\phantom{\nu}\diamond}}{\pm\Upsilon_b\left[m_{\psi}^2-(d-4)b^{(d)\nu\diamond}(b^{(d)})_{\nu}^{\phantom{\nu}\diamond}\right]-(d-4)\Upsilon_b^2}\,,
\end{equation}
and
\begin{align}
\psi_b^{\pm}&=\left\{\pm\Upsilon_b\left[m_{\psi}^2-(d-4)b^{(d)\mu\diamond}(b^{(d)})_{\mu}^{\phantom{\mu}\diamond}\right]-(d-4)\Upsilon_b^2\right\}^{-1} \notag \\
&\phantom{{}={}}\times\left\{(d-3)(\mp\Upsilon_b-b^{(d)\nu\diamond}(b^{(d)})_{\nu}^{\phantom{\nu}\diamond}-m_{\psi}^2)b^{(d)\varrho\sigma\diamond}u_{\varrho}(b^{(d)})_{\sigma}^{\phantom{\sigma}\diamond}\right. \notag \\
&\phantom{{}={}}\hspace{0.4cm}\left.+\,b^{(d)\diamond}\left[b^{(d)\nu\diamond}u_{\nu}+(d-3)b^{(d)\kappa\lambda\diamond}u_{\kappa}p_{\lambda}\right]\right\}\,.
\end{align}
\end{subequations}
Note that we reformulated the denominator in the latter formulas via \eqref{eq:dispersion-equation-b} to eliminate four-momentum components. This procedure has turned out to reduce computation time. Equation~(\ref{eq:equation-lagrangian}) can be solved for the Lagrangian at the cost that it still partially depends on the four-momentum. In total, there are four solutions: two for particles and two for antiparticles. These solutions are very powerful, since they permit us to compute Lagrangians as perturbative series in the controlling coefficients.

The procedure is iterative and for particles it starts with the standard result $L_0^{(d)}=-m_{\psi}\overline{u}$ with $\overline{u}\equiv\sqrt{u^2}$. The latter is obtained from \eqref{eq:equation-lagrangian} in the limit of vanishing Lorentz violation. It is linked to a zeroth-order canonical momentum via $(p_0)_{\mu}\equiv -\partial L_0^{(d)}/\partial u^{\mu}$. Inserting the latter into an appropriate particle solution and keeping all contributions linear in Lorentz violation implies a first-order Lagrangian $L_1^{(d)}$ and another canonical momentum $(p_1)_{\mu}$ valid at first order in Lorentz violation. This iteration can be continued successively to arrive at $L_{q+1}^{(d)}$ valid at $(q+1)$-th order based on the previous iterative results $L_q^{(d)}$ and $(p_q)_{\mu}\equiv -\partial L_q^{(d)}/\partial u^{\mu}$. The corresponding Lagrangians for antiparticles follow from the results for particles via the substitution $m_{\psi}\mapsto -m_{\psi}$. In this case, the iterative procedure starts by inserting $L_0^{(d)}=m_{\psi}\overline{u}$ into an antiparticle solution of \eqref{eq:equation-lagrangian}.

\section{Second-order classical Lagrangians}
\label{sec:third-order-lagrangians}

Experience showed that such perturbative computations to third order in Lorentz violation are feasible for spin-degenerate operators --- independently of their mass dimensions \cite{Schreck:2019mmr}. The situation was quickly revealed to be very different for the spin-nondegenerate cases. First of all, applying the perturbative algorithm described before may require significantly more computation time for these operators. Secondly, while covariant expressions were obtained by a straightforward generalization of particular cases of spin-degenerate coefficients, it turned out to be exceedingly more challenging to accomplish the same for spin-nondegenerate operators.

Considering the controlling coefficients $(K^{(d)})_{\alpha_1\alpha_2\dots}$ of a Lorentz-violating operator of mass dimension $d$, we introduce the following dimensionless tensors of rank $l$ via contractions of the latter with suitable numbers of four-velocities:
\begin{equation}
\label{eq:observer-tensors}
(\tilde{K}^{(d)})_{\alpha_1\dots\alpha_l}\equiv m_{\psi}^{d-4}(K^{(d)})_{\alpha_1\dots\alpha_l\alpha_{l+1}\alpha_{l+2}\dots}\hat{u}^{\alpha_{l+1}}\hat{u}^{\alpha_{l+2}}\dots\,,
\end{equation}
where we employ $\hat{u}^{\alpha}\equiv u^{\alpha}/\overline{u}$. These tensors play an essential role in the construction of covariant forms of Lagrangians. A crucial observation made both in case of the scalar field theory in \cite{Edwards:2018lsn} and for the spin-degenerate fermion operators of the SME \cite{Schreck:2019mmr} is that the tensors that may occur in the Lagrangians have a maximum rank of 2 --- independently of the mass dimension of the operator. The same seems to hold true for the dimension-5 $b$ coefficients. The situation is slightly different for the dimension-5 $H$ coefficients, as the latter are antisymmetric in the first two indices, whereupon a $H^{(5)}_{\mu\nu\varrho\sigma}$ fully contracted with four-velocities is identical to zero: $\tilde{H}^{(5)}=0$. Therefore, tensors of rank 3 at the maximum will be necessary to construct covariant Lagrangians for $\hat{H}_{\mu\nu}$.

In what follows, we will present covariant classical Lagrangians for the dimension-5 $b$ and $H$ coefficients at second order in Lorentz violation. Results will only be given for particles. \ref{sec:procedure-covariantization} provides details on the computational procedure. For the $b$ coefficients we obtained
\begin{widetext}
\begin{align}
\label{eq:lagrangian-perturbative-b}
L^{(5)\pm}_{2,b}&=L_0\Bigg(1\pm \sqrt{(\tilde{b}^{(5)})^2-(\tilde{b}^{(5)})_{\alpha}(\tilde{b}^{(5)})^{\alpha}} \notag \\
&\phantom{{}={}}\hspace{0.5cm}+{}\frac{2(\tilde{b}^{(5)})^4-6(\tilde{b}^{(5)})^2(\tilde{b}^{(5)})_{\alpha}(\tilde{b}^{(5)})^{\alpha}+6(\tilde{b}^{(5)})(\tilde{b}^{(5)})_{\alpha}(\tilde{b}^{(5)})^{\alpha\beta}(\tilde{b}^{(5)})_{\beta}-2(\tilde{b}^{(5)})_{\alpha}(\tilde{b}^{(5)})^{\alpha\beta}(\tilde{b}^{(5)})_{\gamma\beta}(\tilde{b}^{(5)})^{\gamma}}{(\tilde{b}^{(5)})^2-(\tilde{b}^{(5)})_{\alpha}(\tilde{b}^{(5)})^{\alpha}}\Bigg)\,.
\end{align}
\end{widetext}
The latter Lagrangian holds for a totally symmetric choice of $b^{(5)}_{\mu\nu\varrho}$, which corresponds to 20 independent coefficients. The situation is more complicated for the $H$ coefficients. First of all, a slight problem may arise with the notation originally chosen in \cite{Edwards:2018lsn} that we wanted to take over for consistency. Some results may be expressed more conveniently in terms of the dual of $H^{(d)}_{\mu\nu\alpha_1\dots \alpha_{(d-3)}}$, which we will denote by a tilde as usual:
\begin{equation}
(\tilde{H}^{(d)})_{\mu\nu\alpha_1\dots\alpha_{(d-3)}}\equiv \frac{1}{2}\varepsilon_{\mu\nu\varrho\sigma}(H^{(d)})^{\varrho\sigma}_{\phantom{\mu\nu}\alpha_1\dots\alpha_{(d-3)}}\,.
\end{equation}
Note that suitable dimensionless contractions of the dual with $\hat{u}^{\mu}$ are now denoted by a double tilde:
\begin{equation}
(\tilde{\tilde{H}}^{(d)})_{\alpha_1\dots\alpha_l}\equiv m_{\psi}^{d-4}(\tilde{H}^{(d)})_{\alpha_1\dots\alpha_l\alpha_{l+1}\alpha_{l+2}\dots}\hat{u}^{\alpha_{l+1}}\hat{u}^{\alpha_{l+2}}\dots\,.
\end{equation}
Thus, the latter does {\em not} correspond to the dual of the dual in this paper. As a first step, we focus on the minimal dimension-3 coefficients $H^{(3)}_{\mu\nu}$. The Lagrangian for the full set of six independent coefficients has been unknown, so far. Exact results were obtained for the sectors of mixed coefficients $H^{(3)}_{0i}$ and the purely spacelike ones $H^{(3)}_{ij}$ separately. As long as these two sectors do not couple to each other, it holds that $\tilde{Y}^{(3)}\equiv\frac{1}{4}(H^{(3)})_{\mu\nu}(\tilde{H}^{(3)})^{\mu\nu}=0$. The Lagrangian for these cases can be found, e.g., in Eq.~(15) of~\cite{Kostelecky:2010hs}.

Now, for $Y^{(3)}\neq 0$, the algorithm above is employed to obtain a perturbative form of the Lagrangian whose covariantization is much less involved than for the dimension-5 $b$ coefficients. The outcome at third order in Lorentz violation reads
\begin{align}
L^{(3)\pm}_{3,H}&=L_0\left(1\mp \sqrt{-(\tilde{\tilde{H}}^{(3)})_{\alpha}(\tilde{\tilde{H}}^{(3)})^{\alpha}}-\frac{(\tilde{Y}^{(3)})^2}{2(\tilde{\tilde{H}}^{(3)})_{\alpha}(\tilde{\tilde{H}}^{(3)})^{\alpha}}\right. \notag \\
&\phantom{{}={}}\hspace{0.5cm}\left.{}-\frac{(\tilde{Y}^{(3)})^2(\tilde{\tilde{H}}^{(3)})_{\alpha}(\tilde{\tilde{H}}^{(3)})^{\alpha\beta}(\tilde{\tilde{H}}^{(3)})_{\beta\gamma}(\tilde{\tilde{H}}^{(3)})^{\gamma}}{2\left[-(\tilde{\tilde{H}}^{(3)})_{\alpha}(\tilde{\tilde{H}}^{(3)})^{\alpha}\right]^{5/2}}\right)\,.
\end{align}
It is clear that the terms of second and third order must be directly proportional to the quantity $\tilde{Y}^{(3)}$ such that for $\tilde{Y}^{(3)}=0$ the already known Lagrangian in Eq.~(15) of \cite{Kostelecky:2010hs} is reproduced.

A covariant Lagrangian was also obtained successfully for the dimension-5 $H$ coefficients at second order in Lorentz violation. One possibility of writing it up is as follows:
\begin{subequations}
\label{eq:lagrangian-perturbative-H}
\begin{align}
L^{(5)\pm}_{2,H}&=L_0\Bigg(1\mp\sqrt{-(\tilde{\tilde{H}}^{(5)})_{\alpha}(\tilde{\tilde{H}}^{(5)})^{\alpha}}+\frac{\delta L^{(2)}-\delta L^{(2)}_c}{(\tilde{\tilde{H}}^{(5)})_{\alpha}(\tilde{\tilde{H}}^{(5)})^{\alpha}}\Bigg)\,, \displaybreak[0]\\[2ex]
\delta L^{(2)}&=2(S_1^{(5)}-S_5^{(5)}-S_6^{(5)}+S_7^{(5)}+S_8^{(5)}) \notag \\
&\phantom{{}={}}-{}4S_2^{(5)}+S_3^{(5)}-\frac{1}{2}S_4^{(5)}\,, \displaybreak[0]\\[2ex]
\delta L^{(2)}_c&=\frac{1}{2}(\tilde{Y}^{(5)})^2+2(S_7^{(5)}+S_8^{(5)}+S_9^{(5)})\,,
\end{align}
\end{subequations}
with the observer Lorentz scalars expressed in terms of tensors formed from the coefficients $H^{(5)}_{\mu\nu\varrho\sigma}$ (instead of those of the dual operator):
\begin{subequations}
\label{eq:lorentz-scalars-H}
\begin{align}
S_1^{(5)}&=[(\tilde{H}^{(5)})_{\alpha}(\tilde{H}^{(5)})^{\alpha}]^2\,, \displaybreak[0]\\[2ex]
S_2^{(5)}&=(\tilde{H}^{(5)})_{\alpha}(\tilde{H}^{(5)})^{\alpha}\tilde{X}^{(5)}\,, \displaybreak[0]\\[2ex]
S_3^{(5)}&=(\tilde{H}^{(5)})_{\alpha}(\tilde{H}^{(5)})^{\alpha\beta}(\tilde{H}^{(5)})_{\gamma\delta\beta}(\tilde{H}^{(5)})^{\gamma\delta}\,, \displaybreak[0]\\[2ex]
S_4^{(5)}&=(\tilde{H}^{(5)})_{\alpha\beta\gamma}(\tilde{H}^{(5)})^{\alpha\beta}(\tilde{H}^{(5)})_{\delta\epsilon\gamma}(\tilde{H}^{(5)})^{\delta\epsilon}\,, \displaybreak[0]\\[2ex]
S_5^{(5)}&=\hat{u}^{\alpha}(\tilde{H}^{(5)})^{\beta}(\tilde{H}^{(5)})_{\alpha\beta\gamma}(\tilde{H}^{(5)})_{\delta\epsilon}(\tilde{H}^{(5)})^{\delta\epsilon\gamma}\,, \displaybreak[0]\\[2ex]
S_6^{(5)}&=\hat{u}^{\alpha}(\tilde{H}^{(5)})^{\beta}(\tilde{H}^{(5)})_{\alpha\beta\gamma}\hat{u}^{\delta}(\tilde{H}^{(5)})_{\epsilon}(\tilde{H}^{(5)})_{\delta}^{\phantom{\delta}\epsilon\gamma}\,, \displaybreak[0]\\[2ex]
S_7^{(5)}&=(\tilde{H}^{(5)})_{\alpha}(\tilde{H}^{(5)})^{\alpha}\hat{u}^{\beta}(\tilde{H}^{(5)})_{\beta\gamma\delta}(\tilde{H}^{(5)})^{\gamma\delta}\,, \displaybreak[0]\\[2ex]
S_8^{(5)}&=(\tilde{H}^{(5)})^{\alpha}(\tilde{H}^{(5)})_{\beta}\hat{u}^{\gamma}(\tilde{H}^{(5)})_{\gamma\delta\alpha}(\tilde{H}^{(5)})^{\beta\delta}\,, \displaybreak[0]\\[2ex]
S_9^{(5)}&=(\tilde{H}^{(5)})^{\alpha}\hat{u}^{\beta}(\tilde{H}^{(5)})_{\beta\alpha\gamma}(\tilde{H}^{(5)})_{\delta}(\tilde{H}^{(5)})^{\gamma\delta}\,, \displaybreak[0]\\[2ex]
\tilde{X}^{(5)}&=\frac{1}{4}(\tilde{H}^{(5)})_{\alpha\beta}(\tilde{H}^{(5)})^{\alpha\beta}\,, \displaybreak[0]\\[1ex]
\tilde{Y}^{(5)}&=\frac{1}{4}(\tilde{H}^{(5)})_{\mu\nu}(\tilde{\tilde{H}}^{(5)})^{\mu\nu}\,.
\end{align}
\end{subequations}
The latter Lagrangian is valid for a $H^{(5)}_{\mu\nu\varrho\sigma}$ symmetric in its last two indices. The quantities $\tilde{X}^{(5)}$, $\tilde{Y}^{(5)}$ correspond to $\hat{X}$, $\hat{Y}$ of \eqref{eq:definition-X-Y}. The form of $L^{(5)\pm}_{2,H}$ is obviously much more involved than that of $L^{(5)\pm}_{2,b}$.

Several remarks are in order with respect to Eqs.~(\ref{eq:lagrangian-perturbative-b}), (\ref{eq:lagrangian-perturbative-H}). First, for $b^{(5)}_{\mu\nu\varrho}=0$ and $H^{(5)}_{\mu\nu\varrho\sigma}=0$ we obtain the standard result $L_0$, as expected. Second, the term at first order in Lorentz violation for $\hat{b}_{\mu}$ corresponds to that obtained earlier in \cite{Reis:2017ayl} when $d=5$ is inserted. For $\hat{H}_{\mu\nu}$ this is clear, as explained in \ref{sec:peculiarity-H}. Third, the first-order contributions come with distinct signs, which indicates the spin-nondegenerate nature of $\hat{b}_{\mu}$, $\hat{H}_{\mu\nu}$. In contrast, the second-order terms come with a single sign only. In general, modified dispersion relations for spin-nondegenerate operators exhibit an analogous behavior. The degeneracy of the fermion energy with respect to the spin projection is lifted by such operators, which means that, e.g., $\hat{b}_{\mu}$ couples differently to fermions of spin-up in comparison to fermions of spin-down. This different coupling manifests itself via distinct signs in front of contributions that contain odd powers of the Lorentz-violating background. The same is the case when an expression formed of even powers of controlling coefficients occurs inside a square root. The distinct coupling does not play a role, though, for all contributions involving even powers of controlling coefficients.

Fourth, the first-order terms in Lorentz violation are smooth, even when the argument inside the square root vanishes. However, the same does not hold for the second-order term. The denominator can vanish for certain configurations of the controlling coefficients and the four-velocity components, which means that these contributions become singular in these cases (as long as the numerators do not vanish, as well). Hence, for the perturbative expansion to make sense, sufficiently large regions around such singularities in parameter space must be disregarded. Fifth, as the first-order terms contain square roots, their first partial derivatives with respect to the controlling coefficients are not smooth when the expression under the square root vanishes. In particular, the latter holds for $b^{(5)}_{\mu\nu\varrho}=0$ and $H^{(5)}_{\mu\nu\varrho\sigma}=0$, respectively. For the second-order contribution it is the second partial derivatives with respect to the controlling coefficients that are not smooth for certain configurations such as for vanishing controlling coefficients. In general, singularities of classical Lagrangians for spin-nondegenerate operators are attributed to the fact that particle spin, which plays a crucial role for these operators, is a quantum property that cannot be described consistently in the setting of classical physics.

Singularities are known to occur also in the case of the minimal $b$ coefficients. Considering the corresponding Wick-rotated Lagrangian as an algebraic variety, it was explicitly demonstrated that this variety can be desingularized \cite{Colladay:2015wra} in accordance with Hironaka's theorem~\cite{Hironaka:1964}. Finding a suitable desingularization procedure for the cases studied here may be interesting, but is beyond the scope of the paper.

The second-order term for $\hat{H}_{\mu\nu}$ can be decomposed into two contributions. It holds that $\delta L_c^{(2)}=0$ as long as the mixed component operators $\hat{H}_{0i}$ or the purely spacelike ones $\hat{H}_{ij}$ only are considered. When both types of operators are nonzero, $\delta L_c^{(2)}\neq 0$, in general, and must be taken into account. Thus, $\delta L_c^{(2)}$ describes the coupling between the mixed sector and the purely spacelike sector of the dimension-5 $H$ coefficients. In contrast to $\tilde{Y}^{(3)}$ for $d=3$, $\tilde{Y}^{(5)}$ is not sufficient to include all controlling coefficients coupling the two sectors with each other, but additional Lorentz scalars $S_7^{(5)}\dots S_9^{(5)}$ are indispensable.

We also see that 2 is the maximum rank of observer tensors that occur in $L^{(5)\pm}_{2,H}$ such as for the dimension-5 $b$ coefficients and the spin-degenerate operators. Third-rank tensors $(\tilde{H}^{(5)})_{\alpha\beta\gamma}$ play a role, indeed, but their first or second index is always contracted with another $\hat{u}^{\mu}$. Note the antisymmetry in the first two indices, which prohibits us to express such contractions in terms of $(\tilde{H}^{(5)})_{\alpha\beta}$.

Finally, to demonstrate that Eqs.~(\ref{eq:four-velocity-b}), (\ref{eq:four-velocity-H}) can be adopted to the $d$ and $g$ coefficients, we obtained the result for the classical Lagrangian of the minimal, symmetric $d$ coefficients in \eqref{eq:lagrangian-d-2} of \ref{app:minimal-d}. There are some similarities of the latter Lagrangian with \eqref{eq:lagrangian-perturbative-b}, but additional structures occur. Thus, the Lagrangian for the dimension-4 $d$ coefficients at this level is already more involved than that of the dimension-5 $b$ coefficients. This finding clarifies why it has been that challenging to find classical Lagrangians for the minimal $d$ coefficients as opposed to the minimal $b$ coefficients whose exact result was already determined in the very first paper \cite{Kostelecky:2010hs} on classical particle propagation in the SME.

\section{Conclusions and outlook}
\label{sec:conclusions}

In this article we have studied the propagation of a classical, relativistic, pointlike particles in the presence of Lorentz violating-operators of the spin-nondegenerate, nonminimal SME fermion sector. In particular, we applied the algorithm introduced in \cite{Edwards:2018lsn} to the operators $\hat{b}_{\mu}$ and $\hat{H}_{\mu\nu}$ to obtain a quadratic equation in the Lagrangian that can be solved perturbatively in Lorentz violation. By doing so, we were able to derive covariant Lagrangians at second order in Lorentz violation for the totally symmetric dimension-5 $b$ coefficients and the dimension-5 $H$ coefficients symmetric in the last two indices. The computations were involved, but the results obtained demonstrate their feasibility.

It was surprising to find the following generic behavior for~$\hat{b}_{\mu}$:
\begin{align}
L^{(5)\pm}_{2,b}&=L_0\left(1\pm\sqrt{-\mathrm{gram}(\tilde{b},\hat{u})}\right. \notag \\
&\phantom{{}={}}\hspace{0.5cm}\left.{}-\frac{f_b^{(5)}(\tilde{b}^4,\tilde{b}^2(\tilde{b}\cdot\tilde{b}),\tilde{b}(\tilde{b}\cdot\tilde{b}\cdot\tilde{b}),\tilde{b}\cdot\tilde{b}\cdot\tilde{b}\cdot\tilde{b})}{\mathrm{gram}(\tilde{b},\hat{u})}\right)\,,
\end{align}
where we dropped the mass dimension from $\tilde{b}$ for convenience and introduced an intuitive short-hand notation for scalar and matrix products, e.g. $\tilde{b}\cdot\tilde{b}\cdot\tilde{b}\equiv \tilde{b}_{\alpha}\tilde{b}^{\alpha\beta}\tilde{b}_{\beta}$, etc. Furthermore, we employed the Gram determinant $\mathrm{gram}(\tilde{b},\hat{u})\equiv\tilde{b}\cdot\tilde{b}-\tilde{b}^2$ (see the first paper of Ref.~\cite{FinslerPapers}) and $f_b^{(5)}$ is a function characteristic for the dimension-5 $b$ coefficients. It is expected that the behavior for a general mass dimension $d$ is analogous with $f_b^{(5)}$ replaced by $f_b^{(d)}$. The parameters of the linear combination of Lorentz scalars contained in $f^{(d)}$ supposedly depend on the mass dimension $d$, but the overall structure of the function does probably not change. Note that $f_b^{(3)}=0$ such that Eq.~(12) of~\cite{Kostelecky:2010hs}, which is exact for $d=3$, can be reproduced.

Now, the generic form of the Lagrangian for $\hat{H}_{\mu\nu}$ reads
\begin{align}
L^{(5)\pm}_{2,H}&=L_0\left(1\mp\sqrt{-\mathrm{gram}(\tilde{\tilde{H}},\hat{u})}\right. \notag \\
&\phantom{{}={}}\hspace{0.5cm}\left.{}-\frac{f_H^{(5)}((\tilde{H}\cdot\tilde{H})^2,(\tilde{H}\cdot\tilde{H})\tilde{X},\dots)}{\mathrm{gram}(\tilde{\tilde{H}},\hat{u})}\right)\,,
\end{align}
where we again dropped the mass dimension from $\tilde{H}$, $\tilde{\tilde{H}}$ for brevity. Such as before, an intuitive short-hand notation for scalar and matrix products is used as well as the Gram determinant $\mathrm{gram}(\tilde{\tilde{H}},\hat{u})\equiv \tilde{\tilde{H}}\cdot\tilde{\tilde{H}}$. Furthermore, $f_H^{(5)}$ is another function valid for the dimension-5 $H$ coefficients that involves various Lorentz scalars formed from $\tilde{H}$ and $\hat{u}$. It is again reasonable to assume that the generic form of the function $f_H^{(d)}$ is analogous to that of $f_H^{(5)}$, but that the parameters of the linear combinations of Lorentz scalars depend on the mass dimension.

The results of this paper show that perturbative computations of classical Lagrangians for spin-nondegenerate operators are, in principle, feasible despite of them being challenging. Subsets of coefficients that imply simple special cases from Eqs.~(\ref{eq:lagrangian-perturbative-b}), (\ref{eq:lagrangian-perturbative-H}) do not seem to exist.
Computations for the $d$ and $g$ coefficients were not performed explicitly (except for the case of the minimal $d$ coefficients considered in~\ref{app:minimal-d}). However, the basic results stated in Eqs.~(\ref{eq:four-velocity-b}), (\ref{eq:four-velocity-H}) are expected to be taken over conveniently to the case of $\hat{d}_{\mu}$ and $\hat{g}_{\mu\nu}$, respectively, simply by adapting the mass dimension and the number of indices (as described in the paragraph below \eqref{eq:vector-valued-X-Y}). The results presented here may be of interest for mathematicians alike, as they pose the base for constructing Finsler structures beyond those investigated in, e.g., \cite{FinslerPapers,Edwards:2018lsn}. It could also be worthwhile to study cases of Lorentz-invariant operators of the SM effective field theory (see, e.g., Tab.~V in \cite{Kostelecky:2018yfa} and Tab.~XX in~\cite{Kostelecky:2020hbb} in Minkowski spacetime).

Further questions that can still be tackled in the context of classical Lagrangians are as follows: i) What are the results for controlling coefficients that are not totally symmetric? ii) How do the parameters of $f_{b,H}^{(d)}$ depend on the mass dimension $d$? iii) What is the form of higher-order contributions in Lorentz violation for spin-nondegenerate Lorentz violation? iv) How to treat cases with different types of coefficients coupled to each other? However, these problems are rather specific and are probably of lesser interest to the scientific community working on Lorentz violation. Therefore, our conclusion is that with the results of the current paper the problem of classical Lagrangians describing pointlike particles subject to Lorentz violation parameterized by the SME can be considered as solved around 10 years after it was proposed originally in \cite{Kostelecky:2010hs}. A compilation of the Lagrangians obtained in \cite{Kostelecky:2010hs,LagFinslerPapers,Reis:2017ayl,Edwards:2018lsn,Schreck:2019mmr} forms a framework to parameterize Lorentz violation for classical pointlike particles. The latter could be coined the ``point-particle Standard-Model Extension.'' This framework may play a valuable role for theories that lie beyond the hexagon presented in Fig.~2 of the recent paper~\cite{Kostelecky:2020hbb}.

\section*{Acknowledgments}

The authors thank V.A.~Kosteleck\'{y} and B.~Edwards for valuable discussions. M.S. greatly acknowledges support via the grants FAPEMA Universal 01149/17, FAPEMA Universal 00830/19, CNPq Universal 421566/2016-7, and CNPq Produtividade 312201/2018-4. Furthermore, the authors are indebted to CAPES/Finance Code 001.

\appendix
\section{Computation of four-velocities}
\label{sec:computation-four-velocities}

Here we demonstrate explicitly how to derive the four-velocities given in \eqref{eq:four-velocities}. The derivations make extensive use of Eqs.~(\ref{eq:mapping-equations}) as well as of the property that objects like $b^{(d)\diamond}$, $H^{(d)\diamond}$, etc. are positively homogeneous of a certain degree in the four-momentum.

\subsection{Operator $\hat{b}_{\mu}$}

The dispersion equation for the operator $\hat{b}_{\mu}$ reads
\begin{align}
\label{eq:dispersion-equation-b-app}
0&=\left[p^2-b^{(d)\mu\diamond}(b^{(d)})_{\mu}^{\phantom{\mu}\diamond}-m_{\psi}^2\right]^2 \notag \\
&\phantom{{}={}}+4b^{(d)\mu\diamond}(b^{(d)})_{\mu}^{\phantom{\mu}\diamond}p^2-4(b^{(d)\diamond})^2\,.
\end{align}
Its implicit derivative is given by
\begin{align}
0&=\left[p^2-b^{(d)\mu\diamond}(b^{(d)})_{\mu}^{\phantom{\mu}\diamond}-m_{\psi}^2\right]\left(p_0\frac{\partial p_0}{\partial p_j}+p^j-b^{(d)\mu\diamond}\frac{\partial (b^{(d)})_{\mu}^{\phantom{\mu}\diamond}}{\partial p_j}\right) \notag \\
&\phantom{{}={}}+2b^{(d)\mu\diamond}(b^{(d)})_{\mu}^{\phantom{\mu}\diamond}\left(p_0\frac{\partial p_0}{\partial p_j}+p^j\right) \notag \\
&\phantom{{}={}}+2b^{(d)\mu\diamond}\frac{\partial (b^{(d)})_{\mu}^{\phantom{\mu}\diamond}}{\partial p_j}p^2-2b^{(d)\diamond}\frac{\partial b^{(d)\diamond}}{\partial p_j}\,.
\end{align}
Using \eqref{eq:mapping-equation-2-3-4} results in
\begin{align}
0&=\left[p^2-b^{(d)\mu\diamond}(b^{(d)})_{\mu}^{\phantom{\mu}\diamond}-m_{\psi}^2\right]\left(p_0u^j-u^0p^j+u^0b^{(d)\mu\diamond}\frac{\partial (b^{(d)})_{\mu}^{\phantom{\mu}\diamond}}{\partial p_j}\right) \notag \\
&\phantom{{}={}}+2b^{(d)\mu\diamond}(b^{(d)})_{\mu}^{\phantom{\mu}\diamond}\left[p_0u^j-u^0p^j\right] \notag \\
&\phantom{{}={}}-2u^0b^{(d)\mu\diamond}\frac{\partial (b^{(d)})_{\mu}^{\phantom{\mu}\diamond}}{\partial p_j}p^2+2u^0b^{(d)\diamond}\frac{\partial b^{(d)\diamond}}{\partial p_j}\,.
\end{align}
A contraction of the latter derivative with $p_j$ in conjunction with
\begin{subequations}
\begin{align}
p_j\frac{\partial(b^{(d)})_{\mu}^{\phantom{\mu}\diamond}}{\partial p_j}&=(d-3)\left[(b^{(d)})_{\mu}^{\phantom{\mu}\diamond}-\frac{p\cdot u}{u^0}(b^{(d)})_{\mu0}^{\phantom{\mu0}\diamond}\right]\,, \\[2ex]
p_j\frac{\partial b^{(d)\diamond}}{\partial p_j}&=(d-2)\left[b^{(d)\diamond}-\frac{p\cdot u}{u^0}(b^{(d)})_0^{\phantom{0}\diamond}\right]\,,
\end{align}
\end{subequations}
as well as \eqref{eq:mapping-equation-5} implies:
\begin{align}
0&=\left[p^2-b^{(d)\mu\diamond}(b^{(d)})_{\mu}^{\phantom{\mu}\diamond}-m_{\psi}^2\right]\left[p_0(-p_0u^0-L)-u^0p_jp^j\right. \notag \\
&\phantom{{}={}}\left.+\,(d-3)b^{(d)\mu\diamond}\left(u^0(b^{(d)})_{\mu}^{\phantom{\mu}\diamond}+L(b^{(d)})_{0\mu}^{\phantom{0\mu}\diamond}\right)\right] \notag \\
&\phantom{{}={}}+2b^{(d)\mu\diamond}(b^{(d)})_{\mu}^{\phantom{\mu}\diamond}\left[p_0(-p_0u^0-L)-u^0p_jp^j\right] \notag \\
&\phantom{{}={}}-2(d-3)b^{(d)\mu\diamond}\left(u^0(b^{(d)})_{\mu}^{\phantom{\mu}\diamond}+L(b^{(d)})_{0\mu}^{\phantom{0\mu}\diamond}\right)p^2 \notag \\
&\phantom{{}={}}+2(d-2)\left[(b^{(d)\diamond})^2u^0+Lb^{(d)\diamond}(b^{(d)})_0^{\phantom{0}\diamond}\right]\,,
\end{align}
which can be solved for $u^0$. A covariantization provides \eqref{eq:four-velocity-b}.

\subsection{Operator $\hat{H}_{\mu\nu}$}

The dispersion equation has the form
\begin{equation}
0=(p^2-m_{\psi}^2+2\hat{X})^2-8\hat{X}p^2+4H^{(d)\nu\diamond}(H^{(d)})_{\nu}^{\phantom{\nu}\diamond}+4\hat{Y}^2\,,
\end{equation}
with $\hat{X}$ and $\hat{Y}$ defined in \eqref{eq:definition-X-Y}. Its derivative with respect to $p_j$ is given by
\begin{subequations}
\begin{align}
0&=(p^2-m_{\psi}^2+2\hat{X})\left(p_0\frac{\partial p_0}{\partial p_j}+p^j+\frac{\partial\hat{X}}{\partial p_j}\right) \notag \\
&\phantom{{}={}}-2\frac{\partial\hat{X}}{\partial p_j}p^2-4\hat{X}\left(p_0\frac{\partial p_0}{\partial p_j}+p^j\right) \notag \\
&\phantom{{}={}}+2H^{(d)\nu\diamond}\frac{\partial (H^{(d)})_{\nu}^{\phantom{\nu}\diamond}}{\partial p_j}+2\hat{Y}\frac{\partial\hat{Y}}{\partial p_j}\,.
\end{align}
with
\begin{align}
\frac{\partial\hat{X}}{\partial p_j}&=\frac{1}{2}\frac{\partial(H^{(d)})_{\mu\nu}^{\phantom{\mu\nu}\diamond}}{\partial p_j}H^{(d)\mu\nu\diamond}\,, \\[2ex]
\frac{\partial\hat{Y}}{\partial p_j}&=\frac{1}{4}\left(\frac{\partial (H^{(d)})_{\mu\nu}^{\phantom{\mu\nu}\diamond}}{\partial p_j}\tilde{H}^{(d)\mu\nu\diamond}+(\hat{H}^{(d)})_{\mu\nu}^{\phantom{\mu\nu}\diamond}\frac{\partial(\tilde{H}^{(d)})^{\mu\nu\diamond}}{\partial p_j}\right)\,.
\end{align}
\end{subequations}
Inserting \eqref{eq:mapping-equation-2-3-4} leads to
\begin{align}
0&=(p^2-m_{\psi}^2+2\hat{X})\left(p_0u^j-u^0p^j-u^0\frac{\partial\hat{X}}{\partial p_j}\right)+2u^0\frac{\partial\hat{X}}{\partial p_j}p^2 \notag \\
&\phantom{{}={}}-4\hat{X}\left(p_0u^j-u^0p^j\right)-2u^0H^{(d)\nu\diamond}\frac{\partial (H^{(d)})_{\nu}^{\phantom{\nu}\diamond}}{\partial p_j}-2u^0\hat{Y}\frac{\partial\hat{Y}}{\partial p_j}\,.
\end{align}
Multiplication of the latter with $p_j$ gives
\begin{align}
\label{eq:implicit-derivative-H-contracted}
0&=(p^2-m_{\psi}^2+2\hat{X})\left[p_0(-p_0u^0-L)-u^0p_jp^j\right. \notag \\
&\phantom{{}={}}\left.-2(d-3)(u^0\hat{X}+L\hat{X}_0)\right] \notag \\
&\phantom{{}={}}+4(d-3)(u^0\hat{X}+L\hat{X}_0)p^2-4\hat{X}\left[p_0(-p_0u^0-L)-u^0p_jp^j\right] \notag \\
&\phantom{{}={}}-2(d-2)u^0H^{(d)\nu\diamond}(H^{(d)})_{\nu}^{\phantom{\nu}\diamond} \notag \\
&\phantom{{}={}}+L\left[2(d-3)p_{\nu}H^{(d)\nu\varrho 0\diamond}(H^{(d)})_{\varrho}^{\phantom{\varrho}\diamond}+2H^{(d)0\nu\diamond}(H^{(d)})_{\nu}^{\phantom{\nu}\diamond}\right] \notag \\
&\phantom{{}={}}-4(d-3)\hat{Y}(u^0\hat{Y}+L\hat{Y}_0)\,,
\end{align}
where we employed
\begin{subequations}
\begin{align}
\label{eq:homogeneity-relations-H}
p_j\frac{\partial H^{(d)\mu\diamond}}{\partial p_j}&=-\frac{p\cdot u}{u^0}\left[(d-3)H^{(d)\mu\nu 0\diamond}p_{\nu}+H^{(d)\mu 0\diamond}\right] \notag \\
&\phantom{{}={}}+(d-2)H^{(d)\mu\diamond}\,, \\[2ex]
p_j\frac{\partial H^{(d)\mu\nu\diamond}}{\partial p_j}&=(d-3)\left(H^{(d)\mu\nu\diamond}-\frac{p\cdot u}{u^0}H^{(d)\mu\nu 0\diamond}\right)\,, \displaybreak[0]\\[2ex]
p_j\frac{\partial\hat{X}}{\partial p_j}&=\frac{1}{2}p_j\frac{\partial(H^{(d)})_{\mu\nu}^{\phantom{\mu\nu}\diamond}}{\partial p_j}H^{(d)\mu\nu\diamond} \notag \\
&=\frac{d-3}{2}\left((H^{(d)})_{\mu\nu}^{\phantom{\mu\nu}\diamond}-\frac{p\cdot u}{u^0}(H^{(d)})_{\mu\nu 0}^{\phantom{\mu\nu0}\diamond}\right)H^{(d)\mu\nu\diamond} \notag \\
&=2(d-3)\left(\hat{X}-\frac{p\cdot u}{u^0}\hat{X}_0\right)\,, \displaybreak[0]\\[2ex]
p_j\frac{\partial\hat{Y}}{\partial p_j}&=\frac{1}{4}\left(p_j\frac{(H^{(d)})_{\mu\nu}^{\phantom{\mu\nu}\diamond}}{\partial p_j}\tilde{H}^{\mu\nu\diamond}+(H^{(d)})_{\mu\nu}^{\phantom{\mu\nu}\diamond}p_j\frac{\partial\tilde{H}^{\mu\nu\diamond}}{\partial p_j}\right) \notag \\
&=\frac{d-3}{4}\left[\left((H^{(d)})_{\mu\nu}^{\phantom{\mu\nu}\diamond}-\frac{p\cdot u}{u^0}(H^{(d)})_{\mu\nu 0}^{\phantom{\mu\nu 0}\diamond}\right)\tilde{H}^{(d)\mu\nu\diamond}\right. \notag \\
&\phantom{{}={}}\hspace{1cm}\left.+\,(H^{(d)})_{\mu\nu}^{\phantom{\mu\nu}\diamond}\left(H^{(d)\mu\nu\diamond}-\frac{p\cdot u}{u^0}\tilde{H}^{(d)\mu\nu 0\diamond}\right)\right] \notag \\
&=\frac{d-3}{2}\left((H^{(d)})_{\mu\nu}^{\phantom{\mu\nu}\diamond}\tilde{H}^{(d)\mu\nu\diamond}-\frac{p\cdot u}{u^0}(H^{(d)})_{\mu\nu 0}^{\phantom{\mu\nu 0}\diamond}\tilde{H}^{(d)\mu\nu\diamond}\right) \notag \\
&=2(d-3)\left(\hat{Y}-\frac{p\cdot u}{u^0}\hat{Y}_0\right)\,,
\end{align}
\end{subequations}
as well as \eqref{eq:mapping-equation-5} and the dispersion equation~(\ref{eq:dispersion-equation-H}). Here, $\hat{X}_0$ and $\hat{Y}_0$ are the zeroth-order components of the vector-valued quantities defined in \eqref{eq:vector-valued-X-Y}. Note that \eqref{eq:homogeneity-relations-H} has a slightly different form compared to the other relations that follow from the homogeneity of the expressions considered. The reason is that $H^{(d)\mu\nu\diamond}$ is antisymmetric in the first two indices and completely symmetric in the remaining indices only. Finally, \eqref{eq:implicit-derivative-H-contracted} can be solved for $u^0$. Covariantization results in the four-velocity of \eqref{eq:four-velocity-H}.

\section{Covariantization of specific Lagrangians}
\label{sec:procedure-covariantization}

The algorithm described in \secref{sec:results-perturbative-expansion} is usually applied for a certain set of controlling coefficients that are chosen to be the same. For example, considering the isotropic part of the operator $\hat{b}^{(5)}_{\mu}$ that is characterized by the single dimensionless coefficient $x\equiv m_{\psi}b^{(5)000}$ we obtain the following Lagrangian at third order in Lorentz violation:
\begin{subequations}
\label{eq:lagrangian-b-specific-frame}
\begin{align}
L_{b,3}^{\pm}&=L_0\left(1\pm\xi^{(1)}_bx+\xi^{(2)}_bx^2\pm \xi^{(3)}_bx^3+\dots\right)\,, \displaybreak[0]\\[2ex]
\xi^{(1)}_b&=\frac{u_0^2|\mathbf{u}|}{(u^2)^{3/2}}\,, \displaybreak[0]\\[2ex]
\label{eq:lagrangian-b-specific-frame-second-order}
\xi^{(2)}_b&=-2\frac{u_0^2\mathbf{u}^4}{(u^2)^3}\,, \displaybreak[0]\\[2ex]
\xi^{(3)}_b&=4\frac{u_0^2|\mathbf{u}|^5(u_0^2+\mathbf{u}^2)}{(u^2)^{9/2}}\,,
\end{align}
\end{subequations}
where $\mathbf{u}$ is the spatial part of $u^{\mu}$. Note that this result only holds in a single observer frame where all controlling coefficients $b^{(5)}_{\mu\nu\varrho}$ vanish except of $b^{(5)}_{000}$. Analogous results follow for other sets of coefficients. However, it is highly desirable to join all these results to obtain a Lagrangian in covariant form.

It has turned out a formidable task to covariantize the findings that are valid in particular observer frames only. The reason for this is connected to a hitherto unexpected covariant form of the spin-nondegenerate Lagrangians at second order in Lorentz violation. In fact, the covariant second-order contribution was found to be the ratio of an expression at fourth order and an expression at second order in Lorentz violation. Thus, already at second order, suitable observer scalars at fourth order must be taken into account to construct a covariant expression, which renders computations much more challenging than for the spin-degenerate cases. By obtaining a certain number of additional Lagrangians in particular observer frames, we observed that the second-order terms in the denominator correspond to the radicand of the square root that occurs in the contributions at first order in Lorentz violation. Therefore, these terms are of the generic form
\begin{subequations}
\label{eq:generic-second-order-terms}
\begin{equation}
\xi^{(2)}_b=\frac{f_b^{(5)}((\tilde{b}^{(5)})^4,\dots)}{(\tilde{b}^{(5)})^2-(\tilde{b}^{(5)})_{\alpha}(\tilde{b}^{(5)})^{\alpha}}\,,
\end{equation}
for the $b$ coefficients and
\begin{equation}
\xi^{(2)}_H=\frac{f_H^{(5)}((\tilde{H}^{(5)})_{\alpha}(\tilde{H}^{(5)})^{\alpha},\dots)}{(\tilde{\tilde{H}}^{(5)})_{\alpha}(\tilde{\tilde{H}}^{(5)})^{\alpha}}\,,
\end{equation}
\end{subequations}
for the $H$ coefficients. The functions $f_b^{(5)}$ and $f_H^{(5)}$ correspond to linear combinations of observer scalars formed from four copies of the controlling coefficients suitably contracted with four-velocities.

Fortunately, there are certain tools available that turned out to be of great use for constructing observer scalars from a given set of tensors. One of these is the \textit{Mathematica} package \textit{xTras} that is an extension of the package \textit{xTensor} \cite{xTensor:2020}. The latter allows us to define tensors on a manifold endowed with a certain metric that can be interpreted as the Minkowski metric in this case. For the case of the operator $\hat{b}^{(5)}_{\mu}$ we define the second-rank tensor $\tilde{b}^{(5)}_{\mu\nu}\equiv b^{(5)}_{\mu\nu\varrho}u^{\varrho}$ that is assumed to be symmetric in its indices. For $\hat{H}^{(5)}_{\mu\nu}$ we define the third-rank tensor $\tilde{H}^{(5)}_{\mu\nu\varrho}\equiv H^{(5)}_{\mu\nu\varrho\sigma}u^{\sigma}$ that is taken as antisymmetric in the first two indices. \textit{xTras} provides the command \texttt{AllContractions} that is applied on a direct product of four objects $\tilde{b}^{(5)\mu\nu}$ (and $\tilde{H}^{(5)\mu\nu\varrho}$) and a suitable number of four-velocities to form all possible observer scalars. There are 20 possibilities for the case of the dimension-5 $b$ coefficients and 280 for the $H$ coefficients, respectively.

For each specific observer frame these Lorentz scalars are computed and a generic linear combination is formed that is inserted into Eqs.~(\ref{eq:generic-second-order-terms}) where these are then mapped to the second-order terms obtained via the perturbative algorithm (such as~\eqref{eq:lagrangian-b-specific-frame-second-order}). Repeating this procedure for a sufficient number of reference frames, leads to a linear system of equations in the parameters of the generic linear combination. It turned out that the majority of parameters remains free and these are set equal to zero, whereas a very restricted set of parameters has nonzero values. The nonzero parameters for $\hat{b}^{(5)}_{\mu}$ are $\{2,-6,6,-2\}$ and they are multiplied with the Lorentz scalars found in the second-order term of \eqref{eq:lagrangian-perturbative-b}. The number of nonzero parameters is higher for $\hat{H}^{(5)}_{\mu\nu}$, which is not a surprise, as the overall number of coefficients is much higher. They can be read off the second-order term in \eqref{eq:lagrangian-perturbative-H} where they are linked to the Lorentz scalars of \eqref{eq:lorentz-scalars-H}.

\section{Peculiarity for $\hat{H}_{\mu\nu}$ in initial algorithmic step}
\label{sec:peculiarity-H}

A peculiarity arises in the first step of the perturbative method applied to $\hat{H}_{\mu\nu}$. Inserting $(p_0)_{\mu}=m_{\psi}\hat{u}_{\mu}$ into the quadratic equation for $L$, the latter has the following form at first order in the controlling coefficients:
\begin{subequations}
\begin{align}
0&=(1+\upsilon_{\pm})\left(\frac{L}{m_{\psi}}\right)^2+\upsilon_{\pm}\sqrt{u^2}\frac{L}{m_{\psi}}-u^2\,, \\[2ex]
\upsilon_{\pm}&=\frac{2\tilde{X}^{(d)}+(d-2)(\tilde{\tilde{H}}^{(d)})_{\alpha}(\tilde{\tilde{H}}^{(d)})^{\alpha}}{\mp\sqrt{-(\tilde{\tilde{H}}^{(d)})_{\alpha}(\tilde{\tilde{H}}^{(d)})^{\alpha}}}\,.
\end{align}
\end{subequations}
By solving this quadratic equation, one obtains the standard result $L_0=-m_{\psi}\overline{u}$ for particles instead of a first-order Lagrangian such as for $\hat{b}_{\mu}$. Thus, without already knowing the first-order Lagrangian, the perturbative algorithm does not seem to work. Fortunately, first-order results for $\hat{H}_{\mu\nu}$ are already available from the \textit{Ansatz}-based method of Ref.~\cite{Reis:2017ayl}. So we had to employ the findings from the latter paper to obtain the second-order Lagrangian of \eqref{eq:lagrangian-perturbative-H}. Whether or not a physical reason is attributed to this peculiar cancelation at first order in Lorentz violation remains unknown at this moment.

\section{Minimal $d$ coefficients}
\label{app:minimal-d}

For demonstration purposes, we employed the adopted \eqref{eq:four-velocity-b} to obtain a Lagrangian for the minimal $d$ coefficients at second order in Lorentz violation:
\begin{subequations}
\label{eq:lagrangian-d-2}
\begin{align}
L^{(\pm)}_{2,d}&=L_0\left(1\pm \sqrt{(\tilde{d}^{(4)})^2-(\tilde{d}^{(4)})_{\alpha}(\tilde{d}^{(4)})^{\alpha}}\right. \notag \\
&\phantom{{}={}}\hspace{0.7cm}\left.-\frac{f_d^{(4)}}{(\tilde{d}^{(4)})^2-(\tilde{d}^{(4)})_{\alpha}(\tilde{d}^{(4)})^{\alpha}}\right)\,, \displaybreak[0]\\
f_d^{(4)}&=\frac{1}{2}[(\tilde{d}^{(4)})^4-3(\tilde{d}^{(4)})^2(\tilde{d}^{(4)})_{\alpha}(\tilde{d}^{(4)})^{\alpha} \notag \\
&\phantom{{}={}}+4(\tilde{d}^{(4)})(\tilde{d}^{(4)})_{\alpha}(\tilde{d}^{(4)})^{\alpha\beta}(\tilde{d}^{(4)})_{\beta}-[(\tilde{d}^{(4)})_{\alpha}(\tilde{d}^{(4)})^{\alpha}]^2 \notag \\
&\phantom{{}={}}-(\tilde{d}^{(4)})_{\alpha}(\tilde{d}^{(4)})^{\alpha\beta}(\tilde{d}^{(4)})_{\beta\gamma}(\tilde{d}^{(4)})^{\gamma}]\,.
\end{align}
\end{subequations}
The involved structure of the second-order contribution is evident, which explains why it has been challenging to compute any exact results for $d$ in closed form --- even for the minimal coefficients. Note that the Lagrangians for subsets of the minimal $d$ coefficients found in the first two papers of \cite{LagFinslerPapers} cannot be compared to \eqref{eq:lagrangian-d-2}. These are valid for sets of coefficients that are not totally symmetric.


\end{document}